\PassOptionsToPackage{table,dvipsnames}{xcolor}

\documentclass[sigconf]{acmart}
\usepackage{booktabs}

\usepackage{verbatim,graphicx,afterpage,calc,xspace,multirow,booktabs,paralist,wrapfig,listings,fancyhdr,subcaption,array,rotating,hyphenat,textcomp,tikz,multirow}
\usepackage{amsmath,amsfonts,amsthm}
\usepackage{multicol}
\usepackage{paralist}
\usepackage{dblfloatfix}
\usepackage{setspace}
\usepackage{balance}

\renewcommand{\paragraph}[1]{\vspace{0.25em}\noindent{\textbf{#1}}}

\usepackage{color}
\usepackage[bordercolor=white,backgroundcolor=orange!30,linecolor=orange,textsize=scriptsize,prependcaption]{todonotes}

\newif\ifcommenton

\ifcommenton
\newcommand{\sara}[1]{\todo[color=green,caption=Sara]{#1}} 
\newcommand{\bailey}[1]{\todo[color=cyan,caption=Bailey]{#1}} 
\newcommand{\catalina}[1]{\todo[color=lightgray,caption=Catalina]{#1}} 
\newcommand{\ananya}[1]{\todo[color=pink,caption=Ananya]{#1}} 
\else
\newcommand{\sara}[1]{}
\newcommand{\bailey}[1]{}
\newcommand{\catalina}[1]{}
\newcommand{\ananya}[1]{}
\fi

\usepackage{sidecap}
\usepackage{breakurl}
\PassOptionsToPackage{hyphens}{url}
\usepackage{hyperref}
\hypersetup{
    colorlinks = false,
    linkcolor=blue,
    linkbordercolor = {blue}
}

\usepackage{pifont}
\usepackage{fourier-orns}
\definecolor{darkred}{rgb}{.5,0,0}
\definecolor{darkgreen}{rgb}{0,.5,0}
\definecolor{darkyellow}{rgb}{0.95,.6,0.1}

\newcommand{\eberly}{Eberly Center}

\newcommand{\oneless}{$^\dagger$\xspace}
\newcommand{\g}{\cellcolor{lightgray!30}}
\newcolumntype{g}{>{\columncolor{lightgray!30}}c}

\definecolor{linkcolor}{rgb}{0.98,0.97,0.95}
\newcommand{\link}[1]{\textcolor{BlueGreen!90!black}{#1}} 

\usepackage{tcolorbox}
\usepackage{grffile}
\tcbuselibrary{skins}
\definecolor{floralwhite}{rgb}{0.98,0.97,0.95}

\AtBeginDocument{%
  }

\copyrightyear{2023}
\acmYear{2023}
\setcopyright{rightsretained}
\acmConference[SIGCSE 2023]{Proceedings of the 54th ACM Technical Symposium on Computing Science Education V. 1}{March 15--18, 2023}{Toronto, ON, Canada}
\acmBooktitle{Proceedings of the 54th ACM Technical Symposium on Computer Science Education V. 1 (SIGCSE 2023), March 15--18, 2023, Toronto, ON, Canada}
\acmDOI{10.1145/3545945.3569733}
\acmISBN{978-1-4503-9431-4/23/03}

\makeatletter
\gdef\@copyrightpermission{
   \begin{minipage}{0.3\columnwidth}
     \href{https://creativecommons.org/licenses/by-nc/4.0/}{\includegraphics[width=0.90\textwidth]{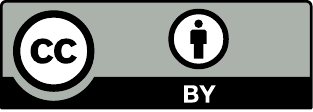}}
   \end{minipage}\hfill
   \begin{minipage}{0.7\columnwidth}
     \href{https://creativecommons.org/licenses/by/4.0/}{This work is licensed under a Creative Commons Attribution International 4.0 License.}
   \end{minipage}
   \vspace{5pt}
}
\makeatother

\title[CS-JEDI: Required DEI Education, by CS PhD Students, for CS PhD Students]{CS-JEDI: Required DEI Education, \textit{by} CS PhD Students, \textit{for} CS PhD Students}
  
\author{Bailey Flanigan}
\affiliation{%
  \institution{Carnegie Mellon University}
  \country{}}
  
\author{Ananya A Joshi}
\affiliation{%
  \institution{Carnegie Mellon University}
  \country{}}

\author{Sara McAllister}
\affiliation{%
  \institution{Carnegie Mellon University}
  \country{}}

\author{Catalina Vajiac}
\affiliation{%
  \institution{Carnegie Mellon University}
  \country{}}

\begin{document}

\begin{abstract}
\vspace{-0.1em}
Computer science (CS) has historically struggled with issues
related to diversity, equity, and inclusion (DEI). Based on how these issues were affecting PhD students in our department (the Carnegie Mellon University CS Department), we identified \textit{required DEI education for PhD students} as a potentially high-impact approach to improving the PhD student experience in our program. Given that no existing curriculum met the desired criteria, we (PhD students)\,---\,alongside many members of the CMU community\,---\,developed and implemented \textit{CS-JEDI: Justice, Equity, Diversity, and Inclusion in Computer Science}. CS-JEDI is a 6-week DEI curriculum that is now taken by all first-year PhD students in our department. 
This paper covers CS-JEDI's motivation and goals;
describes how its evidence-based curriculum is tailored to these goals and to the CS PhD context; and gives a data-driven evaluation of the extent to which CS-JEDI's first offering, in Spring 2022, achieved these goals.

\end{abstract}

\maketitle

\section{Introduction}
\label{sec:intro}
\vspace{-0.1em}
It is well-established that the field of computer science (CS) struggles with pervasive issues related to diversity, equity, and inclusion (DEI). 
At all levels, the field of CS persistently under-represents many groups including women, black and indigenous people of color, queer people, and disabled people~\cite{taulbee_2021, aspray2000recruitment, varma2006making, holloman2021underrepresented,  buzzetto2010unlocking}. 
Underlying this lack of representation are more complex issues\,---\,ranging from field-specific cultural issues to social issues like systemic racism, sexism, and ableism\,---\,
that make it difficult, unwelcome, and sometimes even unsafe for people outside our field's dominant social groups to pursue a career in CS ~\cite{sigarch-women, evans2018evidence, weisshaar2017publish, hammer2020lab, whitney2018increasing, estrada2016improving, dawson2021improving, van2019double}. 
On top of these barriers, PhD students face high rates of mental health issues due to academia's ``dark patterns'' \ananya{someone recently brought up to me that this term in the lab counterculture paper elicits strong feelings of not dark as in bad but dark as in skin tone. Maybe worth not quiting and keeping as (unhealthy patterns) or smthg like this.} \cite{hammer2020lab}, along with ``stress about productivity and self-doubt,...feeling devalued, [issues with] advisor relationships,... difficulties with work-life balance, and feelings of isolation and loneliness'' \cite{satinsky2021systematic}. Despite these struggles, PhD students often do not seek help \cite{satinsky2021systematic}.
While many of these issues affect students of all identities, 
we consider them \textit{DEI issues} because they place disproportionate burdens on community members from marginalized and underrepresented groups \cite{hammer2020lab}. 

As PhD students in Carnegie Mellon's department of CS (CSD), we experienced and witnessed many such DEI issues. 
Based on our experiences, a survey of our peers' experiences, and existing research, we hypothesized that these issues could be mitigated by an introductory DEI course for first-year PhD students. In particular, we envisioned a course engineered to (a) be more comprehensive and self-directed than a standard DEI training, (b) be well-received as a \textit{required} course for \textit{all} students, and (c) connect core DEI topics to the CS PhD context.
While there exist many DEI education programs across industry and academia, none satisfied these criteria:
some are measurably effective but designed to be optional (e.g., \cite{sarch, sfwksp, eberlyCenter, CMUBB, UTTeach}); 
others are required but can be one-size-fits-all or focused on compliance, 
limiting their potential to create lasting knowledge retention or cultural change~\cite{meta2016, higherEdDept, khalid}; 
and the few programs that are both required and more in-depth are not tailored to the CS PhD experience (e.g., \cite{PittMandatory, Pitty1, Ricemand, bp_reqs}).

In light of this gap, we developed our own open-access\footnote{All materials (referenced here in \link{blue}) 
can be downloaded from our \textcolor{blue}{\href{https://www.cs.cmu.edu/~15996/materials.html}{\underline{course website}}}.
} 
DEI curriculum, titled \textit{CS-JEDI: Justice, Equity, Diversity, and Inclusion in Computer Science}.
CS-JEDI is a 6-week introductory DEI course that is, as of Fall 2021, required for all new PhD students in our program. As desired, CS-JEDI is \textit{for} PhD students in that its content and structure are tailored to the CS PhD experience.  
CS-JEDI is also \textit{by} PhD students: it is designed to be mainly PhD student-taught, and it was created primarily by a group of 15 CS PhD students, who contributed their expertise, perspectives, and over 2500 person-hours of work. Beyond this core working group, the curriculum benefited from multiple rounds of detailed input from the CMU Eberly Center for Teaching and Learning, plus many students, staff, faculty, and other experts.

\section{Origin and Motivations of CS-JEDI}
\label{sec:motivation}
\vspace{-0.1em}
CS-JEDI originated in August 2020 when our mounting experiences with DEI issues prompted us to informally survey other PhD students about their experiences in CSD. 
The $\sim$40 anecdotes we heard from $\sim$25 students were consistent with the trends documented in the literature: 
students described interactions with other students\,---\,and sometimes faculty\,---\,involving sexism, racism, xenophobia, homophobia, and harassment. 
Also in accordance with the literature, students reported feeling that their experiences were not understood by others,
struggling to set boundaries and communicate needs in advising relationships, 
feeling unable to intervene in troubling situations, 
experiencing isolation and mental health issues, 
and lacking knowledge of resources.
Students often kept their struggles to themselves: they didn't know they \textit{could} bring them up, lacked the lexicon to describe them, did not know where to seek support, or worried they would be stereotyped or viewed as overly-sensitive for voicing their concerns.
These results suggested to us that CSD could benefit from a course equipping students to (1) actively create inclusive academic environments, 
(2) communicate openly about DEI issues, 
and (3) practice self-care and self-advocacy within academia's inherent power structures.

\paragraph{Why a DEI focus?} 
Being trained primarily in CS and math, incoming CS PhD students may have little prior education on DEI topics. However, in light of the issues identified above, DEI competency has great potential to positively shape students' PhD experiences. Students can benefit from simply being aware of struggles commonly faced in graduate school: if they encounter these struggles, they can name them and know they are not alone. Further, understanding fundamental topics like \textit{intersectionality}, \textit{privilege}, \textit{systemic inequality}, and \textit{stereotype threat} can provide students with language to articulate how these factors are influencing their own experiences, empowering them to advocate for themselves and seek support. Such topics can also increase students' awareness of \textit{others}'\,---\,potentially very different\,---\,experiences, engendering more empathy, inclusive behavior, and support between peers. These benefits can extend to the broader CS community, too:
as future researchers, teachers, and managers, PhD students have great potential to influence our field's culture, 
making it a worthwhile investment to equip them to exert this influence inclusively.

\paragraph{Why a required course?} From the perspective of an institution seeking to prepare students for successful careers, DEI education is sensible: 
in order to secure grants and jobs in academia and industry, applicants are increasingly being required to demonstrate DEI knowledge and contributions to DEI efforts ~\cite{DEIState, NSF}. More importantly, including all students can have much greater community impact: discussion about DEI as a complete cohort can help students find supportive connections they would not have otherwise made. Further, it can build common language between all students on DEI topics and foster a culture in which awareness, open discussion, and accountability are norms. Creating cultural change requires involving all of a community's members \cite{cultureHBR}, but optional DEI courses are unlikely to have such reach, as they are often only attended by
those who are already knowledgeable~\cite{mandVol}. As a result, optional courses may not reach many students who are simply unaware of the potential benefits\,---\,a hypothesis our data will support.  
Moreover, this selection bias is part of a broader pattern of marginalized groups disproportionately contributing labor toward addressing DEI issues \cite{holloman2021underrepresented}\,---\,a trend which is unlikely to be countered with an optional course. In contrast, requiring this work of all students helps to redistribute this labor and establishes through action that creating an inclusive culture is everyone's responsibility.

\section{CS-JEDI Course Goals}
\label{sec:goals}
\vspace{-0.1em}

Now, we describe CS-JEDI's goals in three parts: its \textit{key concepts} (\autoref{sec:concepts});
its \textit{learning objectives}, which apply those concepts in day-to-day life (\autoref{sec:objectives}); and its \textit{high-level teaching goals} (\autoref{sec:course_goals}).

\vspace{-1em}
\subsection{Key Concepts} \label{sec:concepts}
Each week of CS-JEDI focuses on a different key DEI concept. To give students a concrete entry point to these often vast concepts, we frame each one through a \textit{core question}, whose many possible answers touch on some of the concept's central aspects.
For example, Week 2 is on sources of inequality, and its core question is, \textit{How can inequality be perpetuated by a policy/criterion that intends to be neutral to people’s identities?} 
For each week, we define three to four \textit{lenses} through which the core question can be considered. Below, we list each week's key concepts, plus some of their lenses in parentheses. The full list can be found in the \link{Syllabus}.

 \textbf{Week 1:} Introduction and Identity; \textbf{Week 2:} Foundations of Inequality (implicit bias, intersectionality, privilege, oppression); \textbf{Week 3:} Foundations of Identity Safety (stereotype threat, identity safety, structural / interpersonal strategies); \textbf{Week 4:} Intent versus Impact (microaggressions); \textbf{Week 5:} Well-Being in the PhD Program (mental health, boundaries, self-compassion); and \textbf{Week 6:} Allyship and promoting positive change.

\vspace{-0.5em}
\subsection{Learning Objectives} \label{sec:objectives}
CS-JEDI's learning objectives describe skills that students can use to apply the key concepts in their daily lives. These skills can also help address the DEI issues identified in Sections~\ref{sec:intro} and \ref{sec:motivation}.

\vspace{-\topsep}
\begin{enumerate}
\item[\textbf{O1.}] \textbf{Discuss} key concepts openly and inclusively with others.

\item[\textbf{O2.}] \textbf{Name} instances of key concepts in day-to-day scenarios.

\item[\textbf{O3.}] \textbf{Advocate} for and care for oneself using evidence-based practices and campus resources.

\item[\textbf{O4.}] \textbf{Create} an intentionally inclusive space for others using evidence-based practices.

\item[\textbf{O5.}] \textbf{Apply} evidence-based inclusive practices in standard academic contexts (e.g., teaching, admissions, hiring).

\item[\textbf{O6.}] \textbf{Identify} new JEDI topics of interest and self-educate on those topics from multiple perspectives.
\end{enumerate}
\vspace{-\topsep}

\vspace{-0.5em}
\subsection{High-Level Course Goals} \label{sec:course_goals}
CS-JEDI was designed with the goal of being an enjoyable and valuable experience for all students. Toward this aim, it pursues three specific, evidence-based goals:

\paragraph{Goal 1: Create an identity-safe classroom.} 
Colloquially, \textit{identity safety} occurs when students feel they can bring their authentic selves to class. 
More concretely, CS-JEDI aims to treat students' differences as assets \ananya{can cut 'rather than deficits'}rather than deficits, and to ensure that all students can find their experiences reflected in the material---two features of an inclusive class environment according to the \textit{Culturally Responsive Teaching (CRT)} framework \cite{CRT}. 
Since peer interaction is a core aspect of CS-JEDI, meeting this goal means that students' differences are treated as assets by both instructors \textit{and} other students.

\paragraph{Goal 2: Build community among students.} As discussed in Section~\ref{sec:motivation}, PhD students often suffer from isolation and a lack of support. CS-JEDI, being taken by all first-years, offers an opportunity to help students form a supportive community early in graduate school.

\paragraph{Goal 3: Reduce the sense of forced participation.} %
One known pitfall of required DEI education is that it can evoke, for some participants, a sense of forced participation \cite{DEI_Fail}. 
This sense can cause disengagement, create or strengthen anti-DEI convictions \cite{DEI_Fail}, and negatively impact the learning environment.

\vspace{-0.5em}
\section{CS-JEDI Implementation (Spring 2022)} \label{sec:pedagogy}
This section overviews CS-JEDI's main components (\autoref{sec:components}) and then details key features of its design (Sec. \ref{sec:tailor} and \ref{sec:approaches}). In particular, we describe the \textit{Spring 2022} version of the curriculum, which was taught over Zoom (future offerings will be in person).
This version is the product of two years of iterative testing and revision: two previous versions were pilot-tested (once as a requirement), and the curriculum was fully revamped three times based on feedback from students, staff, faculty, and the \eberly.

\vspace{-0.5em}
\subsection{Roadmap of Course Components}
\label{sec:components}
\begin{figure}[H]
    \centering
    \includegraphics[width=0.46\textwidth]{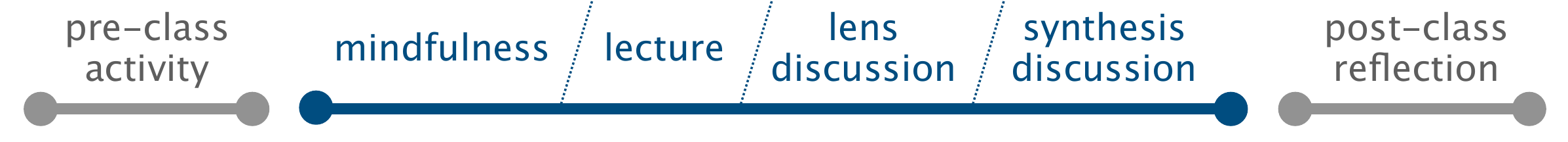}
    \caption{CS-JEDI's weekly schedule. Gray and blue indicate out-of-class and in-class time, respectively.}
    \label{fig:schedule}
\end{figure}

CS-JEDI consists of six, consistently-structured weeks. \autoref{fig:schedule} gives a roadmap of CS-JEDI's weekly curriculum components. CS-JEDI takes <3 hours per week, or 18 hours total, of student time. \\[-0.95em]

\vspace{-0.4em}
\paragraph{Before class.} Students spend 1 hour preparing for class each week through a \link{pre-class activity}.
As a precursor to this activity, all students are assigned one of the current week's lenses. The activity requires each student to research their lens using at least two sources (we provide 10-20 sources per lens; students can also introduce their own) and
document anything\,---\,evidence, opinions, questions\,---\,that they think would be helpful to share with peers.

\paragraph{In class.} Students spend 80 minutes per week in class. Class consists of a 5-minute \textit{mindfulness activity}, a 30-minute \textit{guest lecture} by a domain expert, and then 45 minutes of peer discussion. Peer discussion has two parts: first, students share ideas among others with the same lens in \textit{lens discussions}. 
Then, in \textit{synthesis discussions}, students meet in groups of 4-5, with at least one student per lens, to discuss how their respective lenses inform the core question.

\paragraph{After class.} Students spend 30 minutes on a \link{post-class reflection}, which is a free-form written reflection on that week's material.

\vspace{-0.5em}
\subsection{The key: tailoring to the CS PhD context} \label{sec:tailor} 
CS-JEDI is specifically tailored to CS PhD students\,---\,a feature enabled by the fact that students were its primary architects.
This tailoring allows CS-JEDI to leverage students' common identity and experience as members of a CS PhD program to increase the material's relevance and accessibility to all students at once.

This tailoring comes in two main forms, the first of which is via CS-JEDI's \textit{content}. 
In the opening lecture, the course is motivated with examples of how DEI issues affect members of our department and field and how DEI competency can improve the PhD experience (see \autoref{sec:motivation}). 
Similarly, pre-class activities contain many sources connecting key DEI concepts to CS, academia, and the PhD experience.
For example, these sources describe inclusive research lab cultures \cite{hammer2020lab}, inclusive teaching in mathematical contexts (e.g., \cite{CMUTeach,su2015mathematical, nordell_2022}), 
personal accounts of DEI issues in CS (e.g., \cite{sigarch-women}), 
connections between CS culture and inequities in our field (e.g., \cite{deconstructingCS}), 
and an instructor-curated list of student resources on campus.  
When choosing guest-lecturers, instructors prioritize domain experts from CS-adjacent backgrounds, while also prioritizing those from underrepresented groups. 
All scenarios or thought questions posed to students are similarly specific to the CS context.
Students can also connect a DEI topic to their CS research, or apply CS techniques to learn more about DEI topics, via an optional \link{course project}.
 
We also tailor the course by adopting \textit{structures} and \textit{content framings} that anticipate specific features and challenges of DEI education in the CS PhD context. We motivate and describe these structures and framings throughout the next subsection.

\vspace{-0.5em}
\subsection{Evidence-Based Design Approaches} \label{sec:approaches}
Now, we describe the core evidence-based approaches that CS-JEDI employs toward achieving the goals in Section~\ref{sec:goals}.
Many of these approaches follow one of two frameworks: \textit{Universal Design for Learning} (UDL) \cite{UDL}, which focuses on lowering barriers to learning by giving students choice, and \textit{Culturally Responsive Teaching} (CRT) \cite{CRT}, which recognizes and responds to students' cultural backgrounds, experiences, and identities. 

\paragraph{Preparing students for inclusive participation.} Week 1's class is dedicated to preparing all students\,---\,regardless of their level of experience discussing sensitive topics\,---\,to help create an identity-safe class environment. In week 1, students review the \link{discussion guide}, which provides clear community guidelines, tips for phrasing responses inclusively, and class procedures for accountability and self-advocacy. 
They then practice applying these tools in small groups using \link{provided scenarios}. 
To increase students' awareness of identity-based assumptions they might be making, week 1 also covers \textit{identity} and \textit{intersectionality}, and includes an \link{``identity iceberg'' activity} for exploring how identities can be invisible.

\paragraph{Building instructor and peer support into discussion.} 
Despite this preparation, peer discussion still poses risks to students' identity safety.
In the CS PhD context, we anticipated this risk to be particularly heightened for students from groups that have been historically marginalized in the US, as they are likely to be underrepresented in the course \cite{taulbee_2021}. These students could feel pressured to educate others, or invalidated among peers who share few of their identities or experiences. 
To lessen risks of peer discussion, instructors\,---\,trained in inclusive teaching practices and facilitating difficult dialogues\,---\,can serve as moderators.
Lens groups always have a moderating instructor.
Synthesis groups are too numerous to be constantly moderated, but instructors often check in on groups and are available to moderate if needed or asked.
Because synthesis groups are not always moderated, we build additional, peer support into these groups: first, these groups are fixed, allowing students to build comfort with their group over time.
They are also chosen based on a \link{pre-class survey}, which ensures that students are always placed with someone they have identified as supportive to them, and are not placed with anyone with whom they are uncomfortable.

\paragraph{Incorporating cooperative learning via the Jigsaw Method.} 
The \textit{Jigsaw Method} is a cooperative learning method that, in some contexts, has been found to support learning and reduce prejudice \cite{walker1998academic,aronson2002building}. 
In CS-JEDI's implementation of this method, students individually study their lens
and then share what they learned with others, both within and across lenses. In addition to giving students a chance to contribute to their group's collective knowledge, this approach exposes students to all lens topics in a short time.

\paragraph{Supporting students of all knowledge levels.}  
Students enter CS-JEDI with wide-ranging knowledge levels on all course topics (see ~\autoref{sec:results}). 
To support students when they are new to a topic, we provide a \link{course glossary},
which defines the key concepts and several adjacent terms. 
Additionally, pre-class activities contain clearly-marked introductory sources on every lens.
In framing the content, instructors take care to normalize the difficulty and importance of the subjective questions and qualitative evidence (e.g., firsthand accounts) that are central to studying DEI topics. This may be particularly important for CS PhD students, who are likely more accustomed to quantitative evidence and questions with provable solutions.
Finally, to support students with more incoming knowledge, 
pre-class activities contain many nuanced sources, and students can request reading recommendations from instructors based on their interests. 
Students can also deepen their study by finding their own sources, completing an optional instructor-advised course project, or checking out a book from the \link{class library}. 

\paragraph{Creating space for all perspectives and opinions.} As instructors strongly emphasize, the purpose of CS-JEDI is not to prescribe specific perspectives or opinions; rather, its goal is to help students develop their \textit{own} opinions, based on diverse perspectives and forms of evidence. An essential feature of CS-JEDI is that it is student-taught: this reduces power dynamics between students and instructors, creating space for more open questioning. CS-JEDI is also built around peer discussion, which provides a space for students to\,---\,outside the presence of instructors\,---\,pose their own questions to peers and engage in critical discussion about those questions. Recognizing that any evidence we provide will unavoidably represent limited perspectives, pre-class activities offer numerous and diverse sources per lens and invite students to introduce their own sources. To create space for perspectives from beyond the US context, students can use sources in any language.

\paragraph{Offering diverse modes of learning.} 
A core tool of the two previous approaches was \textit{giving students choice}.
CS-JEDI additionally uses this tool in implementing three tenets of the UDL framework: 
giving choice over \textit{representation} of information, 
\textit{engagement}, 
and \textit{action and expression} \cite{UDL}.
For choice in representation, the sources in pre-class activities include academic research, first-person narratives, and self-reflection tools (e.g., self-surveys). 
These sources also come in diverse formats (e.g., text, audio, video) to suit different learning styles.
For choice in engagement, students can interact with guest lecturers and engage in various ways during peer discussion.
For choice in action and expression,
students can submit notes, diagrams, or any other representation of what they learned in pre-class activities. 
Post-class reflections give students space to practice self-reflection, self-education, and creative expression.

\paragraph{Minimizing mental load on students.}
Time cost and mental load are particular concerns in the PhD context, where students are frequently overworked and strongly incentivized to focus solely on research \cite{hammer2020lab}. 
To limit CS-JEDI's mental load on students, the course is short, has consistent structure week-to-week, and phrases instructions as simply and consistently as possible. 
It is graded pass-fail on the basis of effort, and extensions are given freely upon request. 
To reduce the effort required to engage in class, we begin with a mindfulness exercise, shown to promote engagement~\cite{miralles2021mindfulness}.

\paragraph{Emphasizing empathy, growth, and putting in the work.} 
At the core of CS-JEDI's content is its emphasis on (1) understanding that others may experience the same aspects of our community differently, (2) engaging with the impacts of one's actions and learning from mistakes (rather than focusing on blame \cite{chavez2008beyond}), and (3) taking responsibility for contributing to inclusion in our community. CS-JEDI not only emphasizes these messages, but tries to equip students to enact them: it exposes students to a wide range of others' experiences in our field, teaches intent versus impact, offers students a plethora of sources with which to self-educate in the future, and provides an instructor-curated list of ongoing campus DEI efforts that students can join.
It is important to note that these messages push against cultural norms of our field (and academia more broadly), which emphasizes individual merit and achievement, and in which empathy and service are not always rewarded \cite{hammer2020lab}. The student-taught nature of CS-JEDI is an important asset in overcoming these cultural forces, because it can increase the impact of instructors teaching by example. To model these messages, instructors open class by briefly describing their own processes of putting in the work to learn, unlearn, and advocate for DEI. They also openly invite students to hold them accountable through many low-barrier feedback channels, such as post-class reflections, an \link{anonymous feedback portal}, and a DEI-trained staff contact.

\section{Evaluation Results (Spring 2022)}
\label{sec:results}
\paragraph{Evaluation Methods.} We evaluated the Spring 2022 curriculum with an anonymous \link{course evaluation}, which was designed in collaboration with the \eberly.\footnote{This study was approved by the Institutional Review Board (IRB).} Students had 15 minutes of class time to complete it (but could stay late). The evaluation asked 15 questions, most asking students to rate their agreement with statements according to a five-point \textbf{Likert Scale}:\\[0.4em]
\noindent \includegraphics[width=0.47\textwidth]{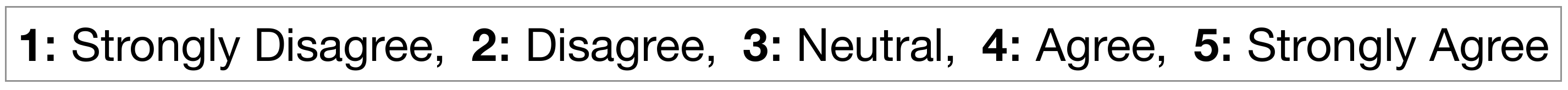}

\paragraph{Demographics.} 30 out of 34 enrolled students submitted the evaluation. We removed two students' responses due to concerns about their reliability (these students reported having \textit{lost} the maximum possible amount of knowledge throughout the course).\footnote{We will, however, consider their short answer responses in our analysis, as they conveyed senses of forced participation.} 
Thus, in the reported data, $n=28$ (or occasionally $n=27$, denoted as \oneless). Per the last question of the survey, 16 students self-identified as international students and 11 self-identified as belonging to a marginalized group (2 declined to answer). We note the possibility that these are underestimates. When most relevant, we will report statistics for these subgroups in addition to aggregate results.

\paragraph{Evaluation of course components.} Before presenting our main results, we summarize students' feedback on CS-JEDI's components. We call a student \textit{positive} toward a component if they \textit{agreed} or \textit{strongly agreed} that it supported their learning. On average, students were solidly positive toward pre-class activities, guest lectures, synthesis group discussions, and the student-taught nature of the course. They were neutral on post-class activities, and slightly below neutral on lens discussions. Most students found the provided instructions clear and felt that the out-of-class work took a reasonable amount of time.

\subsection{Main Results}
\begin{figure}[t]
    \centering
    \includegraphics[width=\columnwidth]{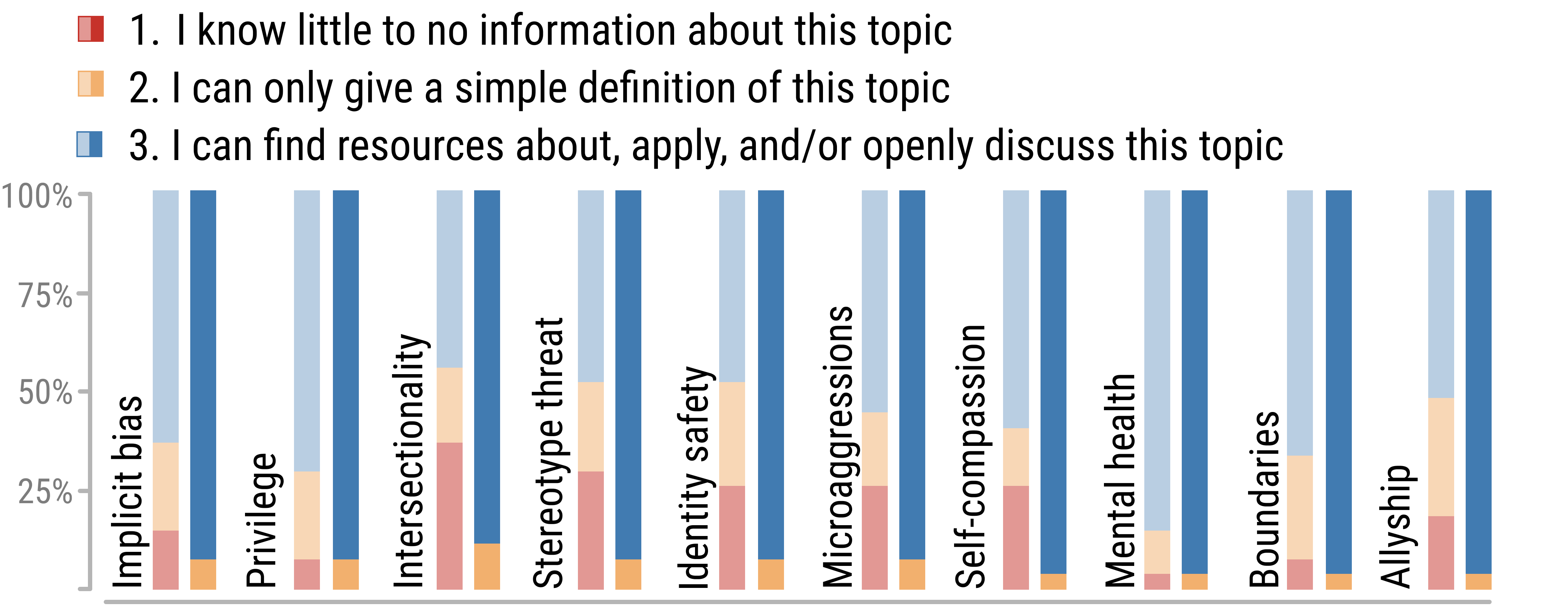}
    \caption{Key concept comprehension pre- (lighter bars) \& post-CS-JEDI.}
    \label{fig:before-after}
\end{figure}

First, CS-JEDI substantially increased students' comprehension of all \textbf{key concepts} (\autoref{sec:concepts}). This is shown in Figure~\ref{fig:before-after}, which compares the distributions of students over three increasing levels of (self-evaluated) comprehension from before to after the course. Despite students' heterogeneous incoming knowledge,  all students reached at least Bin 2\,---\,and \textbf{88\%\,-\,92\%} reached Bin 3, the highest level of comprehension\,---\,on all topics.

\begin{figure}[t]
    \centering
    \includegraphics[width=\columnwidth]{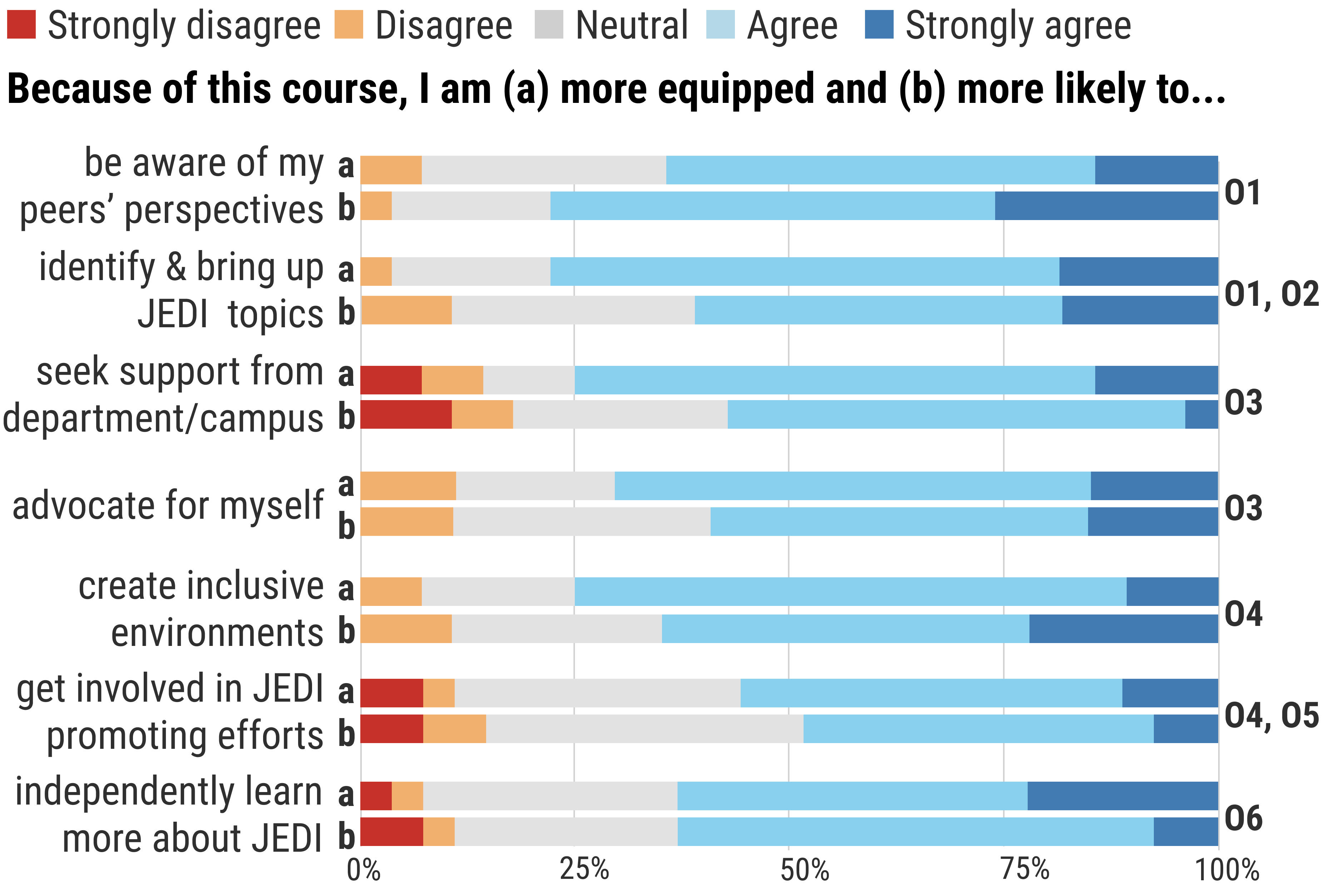}
    \caption{Improvement on learning objectives (O1-O6).}
    \label{fig:community}
\end{figure}

We also find that CS-JEDI helped students develop skills for \textit{applying} this knowledge. As shown in Figure~\ref{fig:community}, \textbf{52\%\,-\,72\%} of students said they were \textit{more equipped}, and \textbf{48\%\,-\,78\%} were \textit{more likely}, to utilize the skills specified in \textbf{learning objectives} O1\,-\,O6 (\autoref{sec:objectives}).

Finally, we evaluate CS-JEDI's achievement of its high level course goals (\autoref{sec:course_goals}) 
through several other questions, mostly summarized in~\autoref{tab:reception}. Note that these questions also directly evaluate many of the specific approaches in Sec. 4.3. Our findings here will show encouraging points of success, but also reveal opportunities for improvement, which we will discuss further in Section \ref{sec:discussion}. 

\textit{Goal 1: Fostering identity safety} is most directly addressed by Question (a), which asked students if they felt they could bring their authentic selves to class. Responses to (a) suggest that a majority of students felt welcome\,---\,and a strong majority did not feel actively \textit{un}welcome\,---\,in class. These results are supported by students' majority positive/neutral responses to questions (b)\,-\,(g), which evaluate specific aspects of identity safety.
Importantly, despite facing potentially heightened risks, marginalized students actually responded far more \textit{positively} than average to these questions. International students, by contrast, often responded less positively.

\begin{table*}[h]
    \centering
    \begin{tabular}{p{7.2cm}|g c |g c |g c}\toprule
     \textbf{Students felt that they...}& \multicolumn{2}{c|}{\textbf{all students } ($n=28$)}  &  \multicolumn{2}{c|}{\textbf{marginalized }($n=11$)} & \multicolumn{2}{c}{\textbf{international } ($n=16$)} \\[0.25em]
     &  Agree+ & Neutral+ & Agree+ & Neutral+ & Agree+ & Neutral+  \\[0.1em]
    \hline
    & \g & & \g & & \g & \\[-0.9em]
    (a) could bring their authentic selves to class & 54\% & 79\%& 64\% & 82\% & 38\%& 63\%\\
    (b) could to alert instructors to concerns \oneless & 59\% & 93\% & 73\% & 91\% & 56\% & 88\% \\
    (c) had the opportunity to voice opinions and questions \oneless & 67\% & 89\% & 91\% & 91\% & 50\% & 81\% \\
    (d) could explore topics relevant to their experiences \oneless & 75\%& 89\% & 91\%& 100\% & 75\%& 88\%\\
    (e) could tailor their learning to the right difficulty & 57\% & 79\% & 82\% & 100\% & 50\% & 75\% \\
    (f) could explore different perspectives / opinions \oneless & 74\% & 89\% & 82\% & 100\% & 69\% & 88\% \\
    (g) could engage with material suited to how they learn & 68\% & 82\% & 91\% & 100\% & 50\% & 75\% \\
    (h) would have taken a course on DEI if not required* & 32\%& 32\% & 64\% & 64\%& 25\%&25\%\\
    (i) are glad they took it \oneless & 63\% & 85\% & 81\% & 91\% & 50\% & 81\% \\
    \bottomrule
\end{tabular}
    \caption{Percentage of students that responded positively to questions related to the pedagogical approaches. Agree+ and Neutral+ correspond to responses $\geq 4$ and $\geq 3$ respectively, per the Likert scale. * denotes a question that did not have a neutral option, and \oneless indicates $n=27$.}
    \label{tab:reception}
\end{table*}

We find that CS-JEDI did promote \textit{Goal 2: build community among students}. In response to a set of questions distinct from those in Table~\ref{tab:reception}, \textbf{63\%} of students reported that after the course, they had more of a \textit{common language} with their peers on DEI topics; \textbf{48\%} said they formed at least one \textit{supportive connection} with another student; \textbf{33\%} were more likely to \textit{ask peers for support}; and \textbf{48\%} of students also felt that they or others would \textit{feel more welcome} in our field if more members of the CS community took this course.

Although two students expressed sentiments of forced participation in free responses (see footnote 3), the majority of responses to all survey questions were neutral or positive, suggesting that most students were receptive to taking CS-JEDI. We evaluate \textit{Goal 3: Reducing the sense of forced participation} more directly with questions (h) and (i), which together test whether students found the course valuable even if they wouldn't have opted into it. We find that the answer is fairly strongly affirmative: even though only 32\% of students would have taken the course had it not been required, 63\% are glad they did. We see that marginalized students were significantly \textit{more} likely to take a DEI course even if not required (64\%), and a higher fraction were glad they did (82\%). International students were much less likely to take the course by choice (25\%), but still, 50\% were glad they took it. Underlying this 50\% was substantial heterogeneity, with 19\% answering ``5'' and 19\% answering ``1''.

\section{Discussion} \label{sec:discussion}
Given the potential pitfalls of required DEI education and the challenges of our context, it was unclear to us at the outset whether a course like CS-JEDI could be successful. We tried to circumvent these pitfalls and challenges through careful course design: we tailored CS-JEDI specifically to our audience, made it peer-taught, and pursued skill-based learning objectives that could be directly applied in students' daily lives. We also involved many students, faculty, and staff in the development process, which helped the course anticipate many context-dependent challenges, and created community investment that has propelled the course toward its establishment in our program's curriculum.

The results presented in this paper\,---\,and our broader experience developing, piloting, and teaching CS-JEDI\,---\,suggest to us that it is possible to design required DEI education programs whose positives outweigh the negatives. CS-JEDI fulfilled many of the promises that motivated us to pursue a required course in the first place: for instance, it not only reached, but \textit{benefited}, many students who were otherwise unlikely to take a DEI-related course during graduate school. These results are consistent with the possibility that required DEI education reaches students who would appreciate it, but are unaware of the potential benefits. Moreover, our finding that marginalized students would have opted into a course like CS-JEDI at far higher rates suggests that had CS-JEDI \textit{not} been required, it could have deepened existing trends of marginalized academic community members undertaking greater service burdens.

Of course, we see these results as only the beginning: the Spring 2022 version we study here was only the \textit{first} of indefinite annual offerings of CS-JEDI. Future iterations will be taught in person, offering many opportunities for new innovations. As we will discuss, the curriculum is designed to respond with agility to student feedback so that it can continue to improve over time. 

\paragraph{Future improvements to the CS-JEDI curriculum.} 
Our results already reveal many opportunities for improving CS-JEDI. At the top of this list, we want the course to do more to strengthen student community, give all students the option to learn at the right level of difficulty (\autoref{tab:reception}, (f)), and make sure there is space for \textit{all} students to bring their authentic selves to class (\autoref{tab:reception}, (a)). One particular community we can better serve in these ways, based on our survey results, is international students, who were overall less positive than average in response to the questions in Table~\ref{tab:reception}. We proceed with the recognition that ``international students'' are an extremely heterogeneous group, as is reflected in our results.

Based on student feedback from throughout the class, we now distill some factors that likely contributed to the negative responses we saw. The material in week 2 is especially difficult, and may have been overwhelming; many course aspects primarily engage with DEI issues in the US context, potentially reducing accessibility or identity safety for students from non-US backgrounds; and students from different cultures may differ in their comfort levels with sharing or submitting their own opinions on sensitive topics.

In response, we identify three planned changes to the curriculum. First, we plan to actively involve students in establishing course motivation, e.g, by beginning class by anonymously crowd-sourcing students' ideas about what marginalization can look like in CS and how greater equity can be achieved. Second, we will front-load more accessible material. To do so, we will re-order the key concepts to first cover well-being in the PhD program (week 5) and identity safety (week 3) before broaching the more challenging (and sometimes more US-centric) topics of inequality, intent versus impact, and allyship (weeks 2, 4, and 6). 
Third, we will try to alleviate pressure on students to share or document their own perspectives and opinions. We will do so by more explicitly encouraging students to share and submit information in \textit{any form}\,---\,e.g., quotes from sources, bullet-point notes, or questions. Further, we will provide synthesis groups with a weekly scenario-based thought question. This can help facilitate discussions centered around concrete experiences, hopefully lessening pressure on students to share their own opinions as a means of generating discussion.

\paragraph{Sustainable and equitable implementation.} 
Surrounding CS-JEDI's curriculum are the institutional processes and resources by which it is sustained over time. These processes were designed with two goals in mind: the curriculum should evolve in response to feedback while remaining high-quality and open-source, and its student instructors should be sufficiently supported, prepared to teach, and compensated. The latter goal is extremely important, given that the PhD student instructors will have high turnover, potentially limited experience teaching, and may face complex power dynamics associated with teaching their peers.

Our \link{sustainable implementation plan} describes how we approach these goals\,---\,how materials are documented and archived over time, and how instructors are selected, compensated, trained, and supported. Additionally, our \link{teaching manual} guides instructors through the process of preparing and teaching the course with detailed instructions. Finally, the curriculum's modular structure permits switching out individual readings, lenses, core questions, and entire key concepts with only self-contained changes to the materials, facilitating low-effort changes in response to feedback.

\paragraph{Adaptability.} This sustainability-promoting infrastructure also makes CS-JEDI's curriculum easily adaptable to new contexts, even outside CS / academia. 
Included in our implementation plan are strategies for implementing CS-JEDI while protecting community members with less institutional power, which are informed by our experiences implementing the course in our own department.

\paragraph{Conclusions.} Although, we (the authors) led various parts of the CS-JEDI initiative, none of us entered our PhD program planning to become so deeply involved in DEI efforts. Like many others who engage in DEI work, we created this course out of frustration and concern about the inequities we saw consistently denying students (including us) an equal opportunity to succeed in our field. We did so \textit{despite} strong incentives not to challenge the status quo.

The fact that so many CS PhD students undertook so much unpaid labor and risk speaks to the urgency of the inequities in CS. These inequities, while pervasive and constantly affecting some, may be invisible to others, even within the same department.\footnote{For example, many faculty in CSD were surprised by the negative experiences students reported in the survey described in Section~\ref{sec:motivation}.} 
This is an argument for educating \textit{everyone} in our field\,---\,students, staff, faculty, and more\,---\,
so that the work required to make our community inclusive can truly be shared among all its members.

\medskip

\noindent\textbf{\textit{Acknowledgements.}} CS-JEDI would not have been possible without the working group of PhD students who offered their time, careful thought, and lived experiences toward the creation of this curriculum: Abhinav Adduri, Alex Wang, Anson Kahng, Josh Williams, Judeth Oden Choi, Pallavi Koppol, Paul Gölz, Samantha Reig, Tabitha E. Lee, Valerie Chen, and Ziv Scully. We are also profoundly thankful for the instrumental curriculum development support provided by Dr.~Alexis Adams of the Eberly Center. We additionally thank Dr.~Chad Hershock, Dr.~Michael Melville, Shernell Smith, and Dr.~Joanna Dickert for their consultation. We thank Professor Zico Kolter for assuming the role of faculty instructor, Deb Cavlovich for constant administrative support, the CSD DEI, SCS DEI, and SCS CS-JEDI steering committees for their support throughout this effort. Innumerable additional faculty, staff, and students from the CMU community also provided invaluable support for CS-JEDI through proofreading, consulting, advocacy, and more, for which we thank them earnestly. 

Bailey Flanigan is supported by a Hertz Foundation fellowship and NSF GRFP, Ananya A Joshi and Catalina Vajiac by NSF GRFP, and Sara McAllister by an NDSEG fellowship.





\bibliographystyle{ACM-Reference-Format}
\balance
\bibliography{acmart.bib}




\end{document}